\documentclass[acmsmall, review=False]{acmart}

\AtBeginDocument{%
  }

\setcopyright{acmlicensed}
\copyrightyear{2024}
\acmYear{2024}
\acmDOI{XXXXXXX.XXXXXXX}

\acmJournal{TOIS}
\acmVolume{37}
\acmNumber{4}
\acmArticle{111}
\acmMonth{8}

\usepackage{graphicx}

\usepackage{color}
\usepackage[noend]{algpseudocode}
\usepackage{algorithmicx,algorithm}
\usepackage{booktabs}
\usepackage{threeparttable}
\usepackage{multirow}
\usepackage{amsmath}
\usepackage{enumitem}

\newlist{Properties}{enumerate}{2}
\setlist[Properties]{label=Prop. \arabic*., font=\textit, itemindent=*}

\begin{document}

\title[\textsc{Co-Matching}: Towards Human-Machine Collaborative Legal Case Matching]{\textsc{Co-Matching}: Towards Human-Machine Collaborative \\Legal Case Matching}

\author{Chen Huang}
\affiliation{%
  \institution{College of Computer Science, Sichuan University}
  \country{Chengdu, China}
}
\email{huangc.scu@gmail.com}

\author{Xinwei Yang}
\email{1531887383@qq.com}
\affiliation{%
  \institution{College of Computer Science, Sichuan University}
  \country{Chengdu, China}
}

\author{Yang Deng}
\affiliation{%
  \institution{School of Computing, National University of Singapore}
  \country{Singapore}}
\email{dengyang17dydy@gmail.com}

\author{Wenqiang Lei}
\affiliation{%
  \institution{College of Computer Science, Sichuan University}
  \country{Chengdu, China}
}
\email{wenqianglei@scu.edu.cn}

\author{JianCheng Lv}
\affiliation{%
  \institution{College of Computer Science, Sichuan University}
  \country{Chengdu, China}
}

\author{Tat-Seng Chua}
\affiliation{%
  \institution{School of Computing, National University of Singapore}
  \country{Singapore}}

\renewcommand{\shortauthors}{Chen et al.}

\begin{abstract}
Recent efforts have aimed to improve AI machines in legal case matching by integrating legal domain knowledge. However, successful legal case matching requires the tacit knowledge of legal practitioners, which is difficult to verbalize and encode into machines. This emphasizes the crucial role of involving legal practitioners in high-stakes legal case matching. To address this, we propose a collaborative matching framework called \textsc{Co-Matching}, which encourages both the machine and the legal practitioner to participate in the matching process, integrating tacit knowledge. Unlike existing methods that rely solely on the machine, \textsc{Co-Matching} allows both the legal practitioner and the machine to determine key sentences and then combine them probabilistically. \textsc{Co-Matching} introduces a method called \textit{ProtoEM} to estimate human decision uncertainty, facilitating the probabilistic combination. Experimental results demonstrate that \textsc{Co-Matching} consistently outperforms existing legal case matching methods, delivering significant performance improvements over human- and machine-based matching in isolation (on average, +5.51\% and +8.71\%, respectively). Further analysis shows that \textsc{Co-Matching} also ensures better human-machine collaboration effectiveness. Our study represents a pioneering effort in human-machine collaboration for the matching task, marking a milestone for future collaborative matching studies.
\end{abstract}


\received{20 February 2007}
\received[revised]{12 March 2009}
\received[accepted]{5 June 2009}

\maketitle

\section{Introduction}
As a crucial aspect of legal research, legal case matching aims at identifying relations between paired legal cases \cite{minocha2015finding, lopez2013link, yu2022explainable, 10285038}, which are long-form documents containing intricate details, such as legal arguments and evidence. Unlike general text matching, legal case matching heavily depends on the tacit knowledge of legal practitioners~\cite{bench2012history}, which encompasses the practical expertise that enables experts to carry out specific actions~\cite{ryle1945knowing}. This tacit knowledge typically includes human intuition, technical skills, and experience that are difficult to articulate and transmit to others~\cite{nonaka1994dynamic, lam2000tacit, gorman2002types}, such as the expertise and work experience in the judicial field. As a result, legal practitioners prioritize investing more effort and exploration to obtain accurate matching results~\cite{shao2021investigating, he2013characterizing, kim2011automatic, mason2006legal}, ensuring the fairness and justice of judicial rulings. However, despite the dedication and willingness of legal practitioners, it is widely recognized that they face challenges in quantitatively measuring the relations between two legal cases, as they struggle to perceive nuances in numerical values~\cite{findling2021computation} and adequately account for complex features~\cite{dellermann201sd9hybrid}. Consequently, \textbf{legal case matching remains a significant challenge for legal practitioners, despite their possession of tacit knowledge}.

Recently, there has been a notable increase in the use of AI models (referred to as machines) to support legal practitioners in legal case matching~\cite{yu2022explainable, bhattacharya2020hier, kumar2011similarity, 10285038}. These machines are appealing because of their potential to accurately discern subtle differences in numerical values and consider complex feature sets when determining relevance and semantic relations, making them the preferred choice for automated legal case matching.
Despite their strengths, emerging evidence suggests that the performance of machines may be hampered when legal knowledge is not adequately integrated into them~\cite{bench2012history, sun2023law, 10.1145/3539618.3591709}. This becomes particularly evident when considering the essential but difficult-to-codify tacit knowledge~\cite{dellermann201sd9hybrid}, defined as skills and experiences that are possessed by people. Such tacit knowledge is derived from a lifetime of experiences across various domains, manifesting in human behaviors~\cite{dellermann201sd9hybrid, ryle1945knowing}. This is challenging to articulate and verbally specify for encoding into the machine. Consequently, a fundamental limitation emerges: \textbf{relying solely on the machine and disregarding the tacit knowledge of legal practitioners struggle to deliver reliable and promising results}.

\begin{figure}
    \centering
    \includegraphics[width=0.85\textwidth]{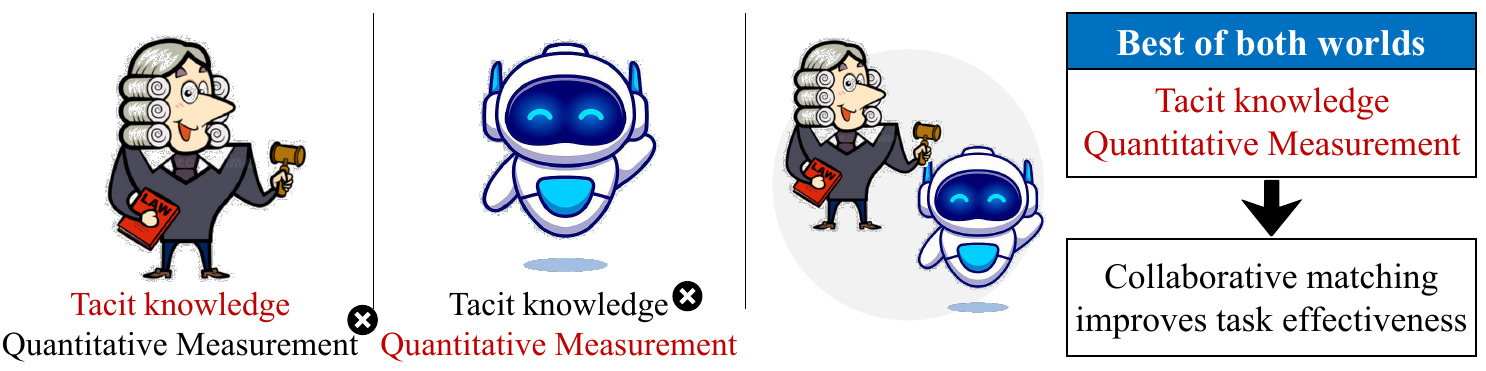}
    \caption{Legal practitioners' tacit knowledge (e.g., work experience) is critical for legal case matching. Human-machine collaborative matching combines the strengths of both the legal practitioner and the machine, leading to enhanced matching results.}
    \label{fig:demo}
\end{figure}

The above observations underscore the vital importance of involving legal practitioners in legal case matching. As shown in Fig. \ref{fig:demo}, we emphasize the significance of harnessing the strengths of both the legal practitioner and the machine. This collaboration results in a human-machine legal case matching process, which we refer to as \textbf{Collaborative Matching}, with the aim of delivering accurate matching results. However, this approach presents two challenges in its implementation.
\begin{itemize}
    \item \textbf{Challenge 1}: \textit{The integration of legal practitioners in the matching process and the combination of their tacit knowledge with the machine is not yet fully understood}. Tacit knowledge, often reflected in human behaviors~\cite{dellermann201sd9hybrid, ryle1945knowing}, necessitates a new collaborative method between legal practitioners and the machine, enabling legal practitioners to utilize their tacit knowledge while overcoming the constraints of manual quantitative measurement. Unlike conversational legal case retrieval~\cite{liu2021conversational, shao2021investigating, liu2022query}, which allows for the expression of intentions and preferences verbally, explicitly verbalizing tacit knowledge for machine encoding is a significant challenge. 
    \item \textbf{Challenge 2}: \textit{The behavior of the legal practitioner typically involves discrete decisions without uncertainty, which presents challenges for human-machine collaboration}. This contrasts with machine outputs, which are typically predictive probabilities. Estimating the uncertainty in human behavior is difficult~\cite{brenner2005modeling, maadi2021collaborative}, particularly after deploying the legal case matching system. This difficulty arises from two issues: the limited behavior data available during system use and the absence of ground truth\footnote{Given the ground truth, it is possible to easily estimate the uncertainty of human behavior, e.g., determining the percentage of errors in human decision-making.} to validate these estimates.
\end{itemize}

\begin{figure*}
    \centering
    \includegraphics[width=0.99\textwidth]{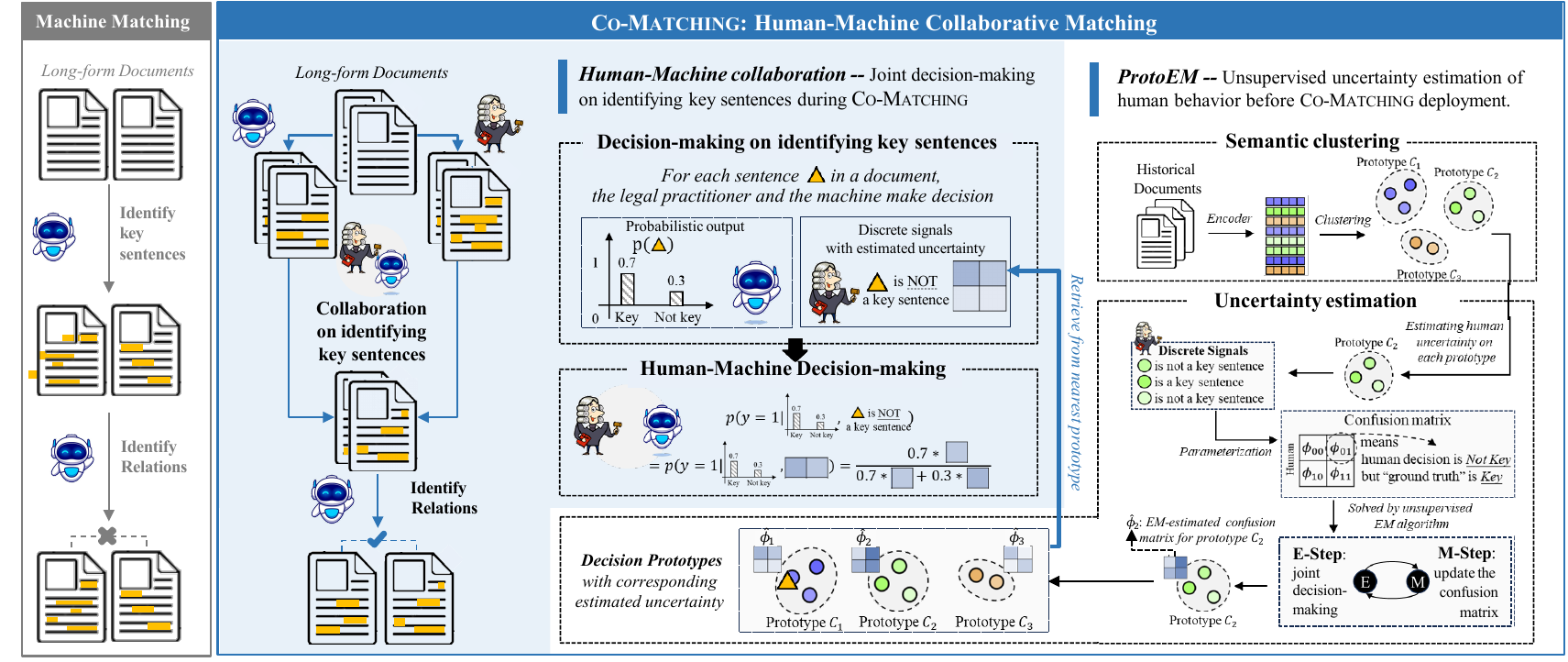}
    \caption{\textsc{Co-Matching} allows both the legal practitioner and the machine to decide on the key sentences and then combine these sentences in a probabilistic manner, while alleviating the limitations of manual quantitative measurement (Challenge 1, marked with a light blue background). \textsc{Co-Matching} introduces the \textit{ProtoEM} to estimate human behavior uncertainty and facilitate the probabilistic decision combination (Challenge 2, marked with a white background on the right side).}
    \label{fig:framework}
\end{figure*}

In light of these challenges, we pioneer work on human-machine collaborative matching and tentatively provide a practical framework, \textsc{Co-Matching} (Collaborative Matching). This framework encourages both the machine and the legal practitioner to participate in the matching process, incorporating tacit knowledge. As illustrated in Fig. \ref{fig:framework}, \textsc{Co-Matching} establishes a novel approach to human-machine collaboration and introduces a new method called \textit{ProtoEM} to estimate human decision uncertainty without assessing the ground truth. 
Specifically, unlike existing methods that rely solely on the machine to identify key sentences in the long-form legal document~\cite{shao2020bert, yu2022explainable, 10285038}, \textsc{Co-Matching} allows both parties to decide on the key sentences and entrusts the machine alone to handle the similarity calculation and search tasks (\textbf{Challenge 1}). By this means, the legal practitioner determines key sentences based on their tacit knowledge (e.g., judgment experience) while avoiding manually handling numerical nuances and complex features. Later, the key sentences selected by both parties are combined in a probabilistic manner to achieve human-machine joint decision-making, which helps eliminate errors in their decisions. 
Additionally, considering the decisions from the legal practitioner are discrete signals, a novel method, \textit{ProtoEM}, is proposed to estimate the uncertainty of the human decision based on the prototype clustering and EM algorithm (\textbf{Challenge 2}). Due to the estimation of uncertainty requiring a large amount of data support, the idea of \textit{ProtoEM} is to estimate the uncertainty of the current human decision by building analogies to the uncertainty of the most relevant decision prototype, where decision prototypes are obtained by clustering all historical decisions of the legal practitioner on each sentence. Given that there is no ground truth, \textit{ProtoEM} estimates the uncertainty of decision prototypes using the Expectation Maximization ({EM}) algorithm, performed on the historical decisions from the corresponding prototype cluster. 
Finally, \textsc{Co-Matching} promotes the joint decision-making of a discrete human decision and a probabilistic machine decision with no ground truth, harnessing the strengths of both the legal practitioner and the mode to promote the machine matching effect.

We evaluate the effectiveness of \textsc{Co-Matching} on two benchmark datasets with manually labeled key sentences for analysis. Experimental results demonstrate that \textsc{Co-Matching} consistently outperforms existing legal case matching methods, emphasizing the importance of legal practitioners and their tacit knowledge. Moreover, in-depth analysis indicates that \textsc{Co-Matching} delivers significant performance improvements over human- and machine-based matching in isolation (on average, +5.51\% and +8.71\%, respectively), indicating the complementarity between the legal practitioner and the machine. Further analysis shows the potential of \textsc{Co-Matching} to collaborate with legal practitioners possessing varying levels of tacit knowledge. Finally, our \textit{ProtoEM} ensures better collaborative efficiency compared to other human-machine cooperation methods. Overall, we pioneer work on human-machine collaborative legal case matching and set a landmark for the future collaborative text matching paradigm. 
To sum up, in this paper, our contributions are as follows:
\begin{itemize}
    \item We emphasize the significance of maintaining the legal practitioner's participation in legal case matching. We underscore the potential for collaboration between the legal practitioner and the machine to work as a team. 
    \item For the first time, we pioneer work on human-machine collaborative matching and tentatively introduce a practical framework, \textsc{Co-Matching}, with the aim of obtaining better performance (Challenge 1).
    \item \textsc{Co-Matching} establishes a novel collaborative approach to inject tacit knowledge and introduces the \textit{ProtoEM} to estimate human decision uncertainty without assessing the ground truth (Challenge 2).
    \item We show experimentally that \textsc{Co-Matching} promotes the matching performance, outperforming both human- and machine-based counterparts in isolation. Additionally, it is suited for legal practitioners of varying levels of tacit knowledge. 
\end{itemize}

\section{Related work}
Different from the legal case retrieval task \cite{10.1145/3539618.3591761}, our research is closely tied to the legal case matching task. Moreover, our research is tied to the human tacit knowledge, human-machine collaboration for legal information processing, and human uncertainty estimation. We will discuss how our work differs from these areas of study.

\subsection{Legal Case Matching}
Legal case matching is a specialized task in text matching that involves identifying relations between paired legal cases to support judges in judicial trial\footnote{This is different from the task of legal case retrieval, which involves finding relevant cases that serve as essential references based on a query case~\cite{10.1145/3539618.3591761}.}~\cite{minocha2015finding, lopez2013link, yu2022explainable, 10285038}. This task requires a high level of specificity and often requires significant effort by legal practitioners to identify semantic relations~\cite{he2013characterizing, kim2011automatic}. The challenges stem from the length of legal documents and the need for strict adherence to legal logic, making this task highly dependent on legal knowledge and practical experience, i.e., the tacit knowledge~\cite{bench2012history}. However, it is widely acknowledged that humans have limitations in quantitatively measuring the relations between two legal cases, as they may struggle to perceive small numerical differences~\cite{findling2021computation} and may not be able to account for a large set of features~\cite{dellermann201sd9hybrid}.
Due to the advancements in deep learning, NLP-based methods have made significant progress in the matching of legal cases. These efforts include developing pre-training techniques to encode large legal corpora~\cite{xiao2021lawformer} and explicitly encoding legal knowledge into machines by, for example, involving precedent citations~\cite{kumar2011similarity, minocha2015finding, bhattacharya2020hier}, decomposing legal issues~\cite{zeng2005knowledge}, establishing an ontological framework for legal problems~\cite{saravanan2009improving}, and considering pro and con rationales in legal documents~\cite{yu2022explainable, 10285038}. However, there is still a substantial amount of tacit legal knowledge that is crucial but difficult to codify for the machine. This highlights the importance of involving legal practitioners in high-stakes legal case matching. For the first time, we harness the potential of both the legal practitioner and the machine initiatives to achieve accurate and comprehensive results. 

\subsection{Tacit Knowledge}
In contrast to formalized, codified, or explicit knowledge, tacit knowledge (or implicit knowledge) is information that is challenging to articulate and is embedded in individual experiences in forms such as motor skills, personal wisdom, intuitions and hunches~\cite{ryle1945knowing}. Consequently, it is difficult to convey to others through writing or verbalization~\cite{nonaka1994dynamic, lam2000tacit, gorman2002types}, and is only evident in human behaviors~\cite{dellermann201sd9hybrid}. While tacit knowledge has been examined in various fields, such as knowledge management \cite{anshari2023optimisation} and organizational behavior \cite{muller2019data}, integrating it into AI models has always posed a challenge \cite{cha2023unlocking, satsangi2019automation, huang2023augmenting}. However, the difficulty in articulating tacit knowledge does not necessarily mean it cannot be formally represented. In this context, the interactive viewpoint of researchers supports the notion that tacit knowledge can be transformed into data \cite{hadjimichael2019toward, fenstermacher2005tyranny} and ontologies \cite{thakker2015padtun}, enabling the transfer of this tacit knowledge. Nevertheless, collecting such data and ontologies presents challenges in achieving thoroughness, accuracy, and generalizability across experiences for various individuals and experts \cite{riedl2016using, piplai2023offline, zajkac2023ground}. Drawing inspiration from a previous study \cite{hadjimichael2019toward} that suggests individuals interact to share their tacit knowledge, we propose a human-machine collaboration paradigm to incorporate the tacit knowledge of legal professionals through mutual interactions, demonstrating its advantages in addressing the legal case matching task.

\subsection{Human-Machine Collaboration for Legal Information Processing}
At present, there is no existing research on the establishment of a collaborative human-machine system for legal case matching. Existing studies have primarily focused on the interactive legal case retrieval setting~\cite{zhai2020interactive}, aiming to simulate the \textit{multi-turn} interactive retrieval process between the user and the retrieval system, taking into account various ways in which a user can convey their preferences through feedback. For example, conversational information retrieval \cite{gao2021advances, zamani2023conversational} enables the machine to gather information that is relevant for understanding user behavior, intentions, and preferences through verbal communication. In the legal domain, researchers have employed legal practitioners as intermediary agents to create an interactive information retrieval system for legal cases~\cite{liu2021conversational}, where the system assists in formulating user queries and examining results. Furthermore, \citet{liu2022query} have developed a real interactive legal case retrieval framework with the ability to generate and suggest queries, leading to higher user satisfaction and success rates. 

However, our work diverges from theirs in two key aspects: 1) We focus on the legal case matching task~\cite{minocha2015finding, lopez2013link, yu2022explainable, 10285038}, as opposed to legal case retrieval~\cite{10.1145/3539618.3591761}. 2) Current methods do not effectively utilize the tacit knowledge of legal practitioners. We argue human initiative should be encouraged, especially when the human is a domain expert. Otherwise, it limits human involvement in the information processing, preventing the full utilization of the strengths of both parties to enhance the performance of legal information processing systems. From a broader perspective, research has explored how humans and computers can work together as a team \cite{wilsoncrossroads}. One popular approach is called \textit{machine-assisted interaction}, where the machine carries out a specific task on behalf of the human (such as finding relevant legal cases), and the human oversees the process and decides whether to accept or reject the machine's output or suggestion. This type of teamwork is seen in existing interactive legal information processing. In this paper, we study the \textit{human-machine collaboration} to harness the potential of both the legal practitioner and the machine capabilities, with the aim of achieving human-machine complementarity \cite{dellermann201sd9hybrid, donahue2022human, bansal2021does}, which means that a combination of human and machine decisions performs better than either humans or machines making decisions on their own. In our case, the proposed \textsc{Co-Matching} involves both the legal practitioner and the machine in the process of identifying key sentences to effectively work together for legal case matching.

\subsection{Human Uncertainty Estimation}
Human decision-making behaviors, which encompass tacit knowledge, are characterized by discrete signals without uncertainty measurement \cite{yang2022optimal, kendall2017uncertainties, cha2021human, oh2020crowd}, making it challenging for machines to evaluate the extent of tacit knowledge involved in decision-making behavior \cite{zhou2019towards}. This underscores the importance of estimating the \textit{epistemic uncertainty} of human decisions \cite{bland2012different}, which pertains to uncertainty stemming from a lack of knowledge or information, i.e., "\textit{we are uncertain because we lack understanding}." Technically, estimating the uncertainty of human decisions is a significant challenge. Previous approaches have favored estimating decision uncertainty using multiple humans and ensemble learning \cite{raghu2019direct}, or by requiring the human to provide soft labels or uncertainty intervals \cite{zhang2021understanding, maadi2021collaborative}. However, these methods are often unreliable, given the high stochasticity in the expression of uncertainty in the human brain \cite{orban2016neural, berkes2011spontaneous}. As a result, other research efforts propose training a human simulator \cite{ma2023should, bourgin2019cognitive} that produces an explicit predictive distribution. Nonetheless, these methods necessitate the collection of extensive human decision data with ground truth of human decision, rendering them impractical for most scenarios. In our case, following the deployment of the legal case matching system, the behavioral data of the current user, i.e., the legal practitioner, is also too limited for uncertainty estimation, and there is no ground truth available. Therefore, it becomes essential to unsupervisedly estimate the uncertainty of legal practitioner behavior to facilitate human-machine collaboration, thus motivating the development of our \textit{ProtoEM} algorithm.

\section{\textsc{Co-Matching}}
The legal practitioner faces challenges in quantitatively measuring the relations between two legal cases. Similarly, the machine struggles to deliver promising performance due to a lack of tacit knowledge. To this end, we pioneer work on human-machine collaborative matching and propose a framework called \textsc{Co-Matching} (cf. Fig.\ref{fig:framework}), including a human-machine collaboration process and the \textit{ProtoEM} to estimate human uncertainty. In Section \ref{over} (\textit{Challenge 1}), we formulate the human-machine collaboration and demonstrate how to inject tacit knowledge without manual quantitative measurement. In Section \ref{compem} (\textit{Challenge 2}), we introduce the proposed method \textit{ProtoEM} and show how to estimate the uncertainty of discrete human decisions without accessing the ground truth. Following previous analysis on legal practitioners \cite{yang2022optimal, 10.1145/3569929}, which indicates that their decision-making behavior deviates from optimality in nuanced ways, the legal practitioner is assumed to be noisy relative to the ground truth.

\subsection{Human-Machine Collaboration (Challenge 1)}
\label{over}
Legal documents are often lengthy, and current methods involve using the machine alone to identify key sentences and filter out irrelevant ones before processing the similarity calculation \cite{shao2020bert, yu2022explainable, 10285038}. Given an off-the-shelf and well-trained legal case matching machine, \textsc{Co-Matching} entrusts both the legal practitioner and the machine to jointly decide on key sentences for each legal document in the input pair, with the machine handling similarity calculation and search tasks. By this means, the legal practitioner uses their tacit knowledge to determine key sentences, such as judgment experience, without manual quantitative measurement. Later, the selected key sentences from both parties are combined in a probabilistic manner to achieve joint decision-making, reducing errors in their decisions. Ultimately, the machine utilizes the combined key sentences in each legal document to identify the relationships between the paired documents.

Formally, we denote $D^K$ as a legal document in the input paired documents containing $K$ sentences\footnote{If the machine handles paragraph-level information such as BERT-PLI~\cite{shao2020bert}, $D^K$ represents a document containing $K$ paragraphs, where $d_i$ represents the $i$-th paragraph.}, where $d_i$ represents the $i$-th sentence. For the convenience of writing, we disregard the superscripts of $D^K$ and simply refer to it as $D$. According to different matching machines, the importance of each sentence can be categorized into different categories, such as the constructive element of crime, the focus of disputes, or not a key sentence. In our paper, we consider $C$ different categories of key sentences. Given an off-the-shelf matching machine, both the legal practitioner and the machine are required to decide the category of the sentence $d_i$, a process referred to as identifying key sentences in this paper.
The legal practitioner's decision on $d_i$, denoted as $H_i$, is a discrete variable with values from the category set, while the machine's decision, denoted as $M_i \in R^C$, is a $C$-dimensional probabilistic vector.

\subsubsection{Machine Decision-Making on Key Sentences}
To improve the matching performance, it is beneficial to identify the key sentences that contain the rationales and legal characteristics and filter out the irrelevant ones. This can be accomplished using various methods such as inverse optimal transport ~\cite{yu2022explainable, 10285038} and semantic modeling~\cite{shao2020bert}. In this paper, we prefer a calibrated machine to avoid the overconfidence problem. This could be achieved in a post-hoc manner by the Temperature Scaling~\cite{guo2017calibration}.
Once the machine makes a decision $M_i$ on each sentence $d_i$, those sentences, together with the human-selected ones, are combined to yield the final key sentences. It's worth mentioning that our focus is not on proposing a better matching machine but rather on ensuring that the legal practitioner works effectively with off-the-shelf matching machines.

\subsubsection{Legal Practitioner Decision-Making on Key Sentences}
The tacit knowledge possessed by legal practitioners is crucial for legal case matching. However, this type of knowledge often manifests in human behavior~\cite{dellermann201sd9hybrid, ryle1945knowing}, making it difficult to verbalize and encode into a machine. To incorporate this tacit knowledge, \textsc{Co-Matching} empowers the legal practitioner to make a decision $H_i$ on identifying the importance of each sentence $d_i$. This allows the legal practitioner to participate deeply in the matching task and eliminates the need for manual numeric calculation. Additionally, it's important to note that 1) the time and cognitive load introduced by the human decision-making process are insignificant. Previous studies have shown that legal practitioners prefer to invest more effort in using domain-specific knowledge to enhance legal-related tasks~\cite{shao2021investigating, he2013characterizing}. 2) While the decision-making of legal practitioners may not be entirely accurate~\cite{yang2022optimal, 10.1145/3569929}, legal practitioners are at least striving to perform tasks optimally in the process of judicial judgment and legal case matching, rather than making completely random decisions regarding the key sentences.

\subsubsection{Human-Machine Collaborative Decision-Making}
Given the input sentence $d_i \in D$, the machine decision $M_i$, and the legal practitioner's decision $H_i$, \textsc{Co-Matching} aims to combine the $H_i$ and $M_i$ to achieve human-machine joint decision-making regarding the key sentences. While the simplest method for this is ensemble learning, such as weighted majority voting, it is not effective in our case. Only two parties are involved in the collaborative matching task, and ensemble learning cannot outperform the best decisions of both parties. Inspired by previous studies~\cite{xu1992methods, kuncheva2014combining, kerrigan2021combining}, \textsc{Co-Matching} combines $H_i$ and $M_i$ in a probabilistic manner, which fits the discrete decision of the legal practitioner into the confusion matrix.

Formally, we denote the human-machine joint decision on sentence $d_i$ as $p(y_i|M_i, H_i)$, where $y_i$ represents the \textit{ground truth} decision on $i$-th sentence. Assuming that the legal practitioner decision $H_i$ and the machine decision $M_i$ are conditionally independent given $y_i$, the joint decision $p(y_i|M_i, H_i)$ follows the Bayesian rule.
\begin{equation}
\label{ddd}
    p(y_i|M_i, H_i)\propto p(H_i|M_i, y_i)p(y_i|M_i)
    \propto p(H_i|y_i)p(y_i|M_i),
\end{equation}
where the term $p(y_i|M_i)$ represents the predictive distribution of the matching machine, while $p(H_i|y_i)$ captures the uncertainty in the legal practitioner's decision at the class level. Technically, this uncertainty can be represented as a confusion matrix~\cite{xu1992methods, kuncheva2014combining} denoted as $\phi_i$, where each element $[\phi_i]_{jk}=p(H_i=j|y_i=k)$ can be calculated by assessing the inconsistency rate between the ground truth $y_i$ and $H_i$ on sentence $d_i$. In this scenario, the joint decision $p(y_i|M_i, H_i)$ can be re-expressed in a closed form, where the denominator term is used for re-normalization:
\begin{equation}\label{ddddd}
p(y_i=j|M_i, H_i=k)=\frac{[\phi_i]_{kj} M_{ij}}{\sum_{q=1}^C [\phi_i]_{kq} M_{iq}}.
\end{equation}
Here, $M_{ij}$ represents the value of the $j$-th dimension of $M_i$. By this means, \textsc{Co-Matching} establishes the human-machine collaboration that introduces tacit knowledge via human decision-making on key sentences and combines them, together with the machine decisions, to yield the joint decisions on the key sentences. However, despite the closed-form joint decision $p(y_i=j|M_i, H_i=k)$, the challenge lies in estimating human uncertainty, which is parameterized by $\phi_i$. Specifically, when a legal practitioner uses the system and inputs a new legal document pair with each document is represented as $D$, 1) the ground truth ${y}_i$ for each sentence ${d}_i \in {D}$ is unknown. 2) Furthermore, there are not enough samples to estimate the uncertainty ${\phi}_i$ of each human decision made for each sentence ${d}_i \in {D}$. This limitation impedes the coordination capability of \textsc{Co-Matching} to efficiently adapt to various legal users after deployment. We address how to solve this problem in Section \ref{compem}.

\subsubsection{Identifying Relations Between Paired Legal Documents}
\textsc{Co-Matching} involves the machine to identify the semantic relations between the input legal documents using selected key sentences. In contrast to previous methods use machine-selected sentences \cite{yu2022explainable, 10285038, shao2020bert}, \textsc{Co-Matching} uses sentences chosen by both parties, i.e., the joint decision-making $p(y_i=j|M_i, H_i=k)$. Subsequently, the machine predict the most semantic relations as per standard procedure.

\subsection{\textit{ProtoEM}: Estimating Human Uncertainty  (Challenge 2)}
\label{compem}
Human decision-making behaviors, while containing tacit knowledge, involve discrete signals~\cite{yang2022optimal}, which hinders the achievement of joint decision-making. Estimating the uncertainty of human decisions is a significant challenge. Previous efforts prefer training a human simulator \cite{ma2023should, bourgin2019cognitive} that outputs an explicit predictive distribution. However, these methods require ground truth and extensive data, making them unsuitable for our case. To address this, our \textit{ProtoEM} leverages a large set of historical documents (e.g., the training data of the matching machine), which records historical decision-making behaviors of all legal practitioners, and employs the EM algorithm to estimate the uncertainty of current human decisions, thus bypassing the aforementioned problems. Specifically, \textit{ProtoEM} first clusters all sentences in historical documents to obtain $P$ prototypes and then estimates human decision uncertainty for each prototype. It's important to note that this entire process can be completed before the deployment of \textsc{Co-Matching}. When the legal practitioner uses \textsc{Co-Matching}, it obtains the uncertainty of the current human decision by drawing analogies to the uncertainty of the most relevant historical prototype without further calculation.

\begin{algorithm}[t]
    \caption{{Pseudo-code of \textsc{Co-Matching}}}
    \label{alg:algorithm1}
\begin{algorithmic}
\State\textbf{Input}: Historical documents $\mathbf{D}$, a legal case pair $Ds$, one matching machine $M$, one legal practitioner $H$, and the number of EM iterations $\#EM$ 
\State\textbf{Output}: Relevant legal case
\end{algorithmic}
\begin{algorithmic}[1] 
        \State \%\%\% \textit{Before deploying \textsc{Co-Matching}.} \%\%\% 
        \State Cluster all sentences $d_i \in \mathbf{D}$ to yield prototypes.
        \State Set $i=0$.
        \For{$i < \#EM$}
        \State $i += 1$.
        \State E-Step: Obtain joint decision by $p(\hat{y}_i|\hat{M}_i, \hat{H}_i, \hat{\phi}^{(t-1)}_j)$.
        \State M-Step: Together with $p(\hat{y}_i|\hat{M}_i, \hat{H}_i, \hat{\phi}^{(t-1)}_j)$, update the confusion matrix by $[\hat{\phi}_j^{(t)}]_{kq} = \frac{\sum_i \sum_k \mathbf{1}[\hat{H}_i=q] \cdot p(\hat{y}_i=k|\hat{M}_i, \hat{H}_i, \hat{\phi}^{(t-1)})}{\sum_i \sum_k p(\hat{y}_i=k|\hat{M}_i, \hat{H}_i, \hat{\phi}^{(t-1)})}$
        \EndFor
        \vspace{3mm}
        \State \%\%\% \textit{Use \textsc{Co-Matching} in practice.} \%\%\% 
        \For{document $D \in Ds$}
        \For{sentence $d_i \in D$}
            \State Machine decision-making on key sentence $M_i$. 
            \State Legal practitioner decision-making on key sentence $H_i$.
            \State Obtain $\phi_i = \hat{\phi}_j$, where $j=\arg\min_{k} \| d_i - \overline{d}_{p_k} \|$.
            \State Human-machine decision-making on key sentence by the equation $p(y_i=j|M_i, H_i=k)=\frac{[\phi_i]_{kj} M_{ij}}{\sum_{q=1}^C [\phi_i]_{kq} M_{iq}}$.
        \EndFor
        \EndFor
        \State The machine uses the selected key sentences to identify semantic relations between paired legal documents as per standard procedure.
\end{algorithmic}
\end{algorithm}

\subsubsection{Semantic clustering}
Formally, we have a set of historical documents denoted as $\mathbf{D}={\hat{D}_1,...,\hat{D}_T}$. For each sentence $\hat{d}i \in \mathbf{D}$, we obtain its semantic embedding from a legal-tailored BERT\cite{zhong2019openclap}, using the concatenation of $[\hat{d}i, \hat{d}{\neg i}]$ as input, where $\hat{d}{\neg i}$ represents the surrounding context sentences of length $L$. Subsequently, we use the K-Means algorithm to cluster all embeddings into $P$ prototypes. Each prototype captures the similar decision-making behavior of legal practitioners in nuanced semantic contexts.

\subsubsection{Uncertainty Estimation via EM}
\label{emduwe}
\textit{ProtoEM} estimates the uncertainty of decisions for each prototype. For each sentence $\hat{d}_i$ from the $j$-th decision prototype $p_j$ and its associated human decision $\hat{H}_i$ and machine decision $\hat{M}_i$, \textit{ProtoEM} aims to estimate the corresponding decision uncertainty at a class level, parameterized by the confusion matrix $\hat{\phi}_j$. In the absence of ground truth, \textit{ProtoEM} turns to the EM algorithm, an optimization method that iteratively provides a solution for maximum likelihood estimation with latent variables (i.e., unknown ground truth in our case) using a batch of available data (i.e., prototype $p_j$). Intuitively, in the E-step, we use $\hat{\phi}_j$ at hand to calculate the joint human-machine decision. In the M-step, we use the human-machine joint decision as the ground truth to update $\hat{\phi}_j$ for prototype $p_j$. Formally, at round $t$, our objective function is expressed as follows:
\begin{equation}
\label{what}
    \hat{\phi}^{(t)}_j = \arg\max_{\hat{\phi}_j} \sum_{\hat{y}_i} \sum_{\hat{H}_i} p(\hat{y}_i|\hat{M}_i, \hat{H}_i, \hat{\phi}_j^{(t-1)}) \log p(\hat{y}_i, \hat{M}_i, \hat{H}_i | \hat{\phi}_j)
\end{equation}
Since the legal practitioner and the machine make independent decisions on key sentences, and the confusion matrix $\hat{\phi}_j$ captures class-level information, the term
$p(\hat{y}_i, \hat{M}_i, \hat{H}_i | \hat{\phi}_j)$ could be decomposed into $p(\hat{M}_i)p(\hat{H}_i)p(\hat{y}_i | \hat{M}_i, \hat{H}_i, \hat{\phi}_j)$ via the Total Probability Theorem. By plugging this into the problem (\ref{what}) and omitting constant terms that do not depend on $\hat{\phi}_j$, we obtain the following equation:
\begin{equation}
\label{what2}
    \hat{\phi}^{(t)}_j = \arg\max_{\hat{\phi}_j} \sum_{\hat{y}_i} \sum_{\hat{H}_i} p(\hat{y}_i|\hat{M}_i, \hat{H}_i, \hat{\phi}^{(t-1)}_j) \log p(\hat{H}_i |\hat{y}_i, \hat{\phi}_j).
\end{equation}
Note that the first term $p(\hat{y}_i|\hat{M}_i, \hat{H}_i, \hat{\phi}^{(t-1)}_j)$ is the joint decision made by both the legal practitioner and the machine, and it is a constant that can be computed in the E-step. As a result, the optimization problem (\ref{what2}) could be easily solved in a closed form. Specifically, the element $[\hat{\phi}_j]_{kq}$ of the confusion matrix for the $j$-th prototype is determined by the following equation
\begin{equation}
[\hat{\phi}_j^{(t)}]_{kq} = \frac{\sum_i \sum_k \mathbf{1}[\hat{H}_i=q] \cdot p(\hat{y}_i=k|\hat{M}_i, \hat{H}_i, \hat{\phi}^{(t-1)})}{\sum_i \sum_k p(\hat{y}_i=k|\hat{M}_i, \hat{H}_i, \hat{\phi}^{(t-1)})},
\end{equation}
where $\mathbf{1}[\cdot]$ is an indicator function. Using this approach, we create a confusion matrix $\hat{\phi}_j$ for each prototype $p_j$, which quantifies the uncertainty of legal practitioners in similar decision-making behaviors without requiring access to the ground truth. 

\subsubsection{Human-machine collaboration based on ProtoEM}
\label{asdrfwef}
We summarize the usage of \textsc{Co-Matching} in the pseudo-code (cf. Algorithm \ref{alg:algorithm1}). Before deploying \textsc{Co-Matching} for practical use, we run \textit{ProtoEM} using historical documents (e.g., the training data of the legal case matching model), which record historical decision-making behaviors of all legal practitioners, and obtain the decision prototypes together with their estimated human uncertainty. After the deployment, for each sentence $d_i$ of the input document $D$, we quantify the uncertainty of the human decision $H_i$ by building analogies to the uncertainty of the most relevant decision prototype. Formally, the uncertainty $\phi_i$ for $H_i$ is determined by the following equation, where $\overline{d}_{p_k}$ is the averaged sentence embedding over prototype $p_k$.
\begin{equation}
    \phi_i = \hat{\phi}_j \text{\quad\quad where } j=\arg\min_{k} \| d_i - \overline{d}_{p_k} \|
\end{equation}

\section{Experiments}
We call attention to the importance of preserving the legal practitioner’s involvement in legal case matching. Closely revolving around this, we conduct extensive experiments to assess the effectiveness of \textsc{Co-Matching}. Given off-the-shelf legal case matching machines, we evaluate if \textsc{Co-Matching} is more desirable to obtain more accurate and comprehensive results, compared to baselines (cf. Section \ref{dyem}). Furthermore, we comprehensively analyze the advantages of \textsc{Co-Matching} and uncover its characteristics (cf. Section \ref{dyem2}).

\subsection{Experiment Setup}
\label{setup}
\textbf{Baseline}. Based on our understanding, there is currently no method specifically designed to incorporate the tacit knowledge of legal practitioners into the machine and establish collaborative matching. To assess the effectiveness of \textsc{Co-Matching}, we include commonly used and SOTA legal case matching methods as baselines. 
Of these, \textit{BERT-PLI}, \textit{IOT-Match}, and \textit{GEIOT-Match} are our primary baselines, as they identify key sentences in the input and filter out irrelevant ones to improve the matching task. Additionally, we include \textit{GPT-3.5-16K-turbo} in our experiments, which is a highly effective baseline among legal-specific LLMs~\cite{fei2023lawbench}.
\begin{itemize}
    \item \textbf{Sentence-BERT} \cite{reimers2019sentence} uses legal-specific BERT to encode legal documents and employs a MLP for legal case matching. Note that Sentence-BERT is a commonly used baseline in legal case matching task \cite{yu2022explainable, 10285038}.
    \item \textbf{Lawformer} \cite{xiao2021lawformer} utilizes a legal-specific Longformer \cite{beltagy2020longformer}, which is a pre-trained language model, to encode long legal documents. 
    \item \textbf{Thematic Similarity} \cite{bhattacharya2020methods} computes the paragraph-level similarities. It then uses the maximum or average similarities to make an overall prediction of the matching between the two cases. Note that both the Thematic Similarity$_{max}$ and Thematic Similarity$_{avg}$ are utilized in our experiments.  
    \item \textbf{BERT-PLI} \cite{shao2020bert} uses BERT to capture semantic relations at the paragraph level and then combines them using MaxPooling and RNN to predict the relevance of two legal cases.
    \item \textbf{IOT-Match} \cite{yu2022explainable} utilizes the inverse optimal transport to identify key sentences, avoiding the negative impact of noise sentences on the matching results. Then, IOT-Match learns to conducts matching only based on the extracted sentences via MLP.
    \item \textbf{GEIOT-Match} \cite{10285038} constructs a heterogeneous graph to explicitly represent the legal cases and their relations. It is an improved method over IOT-Match with a graph-based inverse optimal transport module, achieving SOTA on our evaluation datasets.
    \item \textbf{GPT-3.5-16K-turbo} \cite{chat-gptt}. Recent benchmark \cite{fei2023lawbench, dai2023laiw} show that existing legal-specific LLMs are not as effective as GPT-3.5-16K-turbo\footnote{We omit the GPT-4 due to financial constraints. Legal documents are very lengthy.}. Additionally, legal case matching requires processing very long documents, which surpass the length limit for most LLMs \cite{fei2023lawbench}. Due to the length limit of GPT-3.5-16K-turbo, we only consider the zero-shot setting.
\end{itemize}
To analyze \textsc{Co-Matching}'s advantages and uncover its characteristics, we consider the following ablation baselines and alternatives to achieve human-machine collaboration:
\begin{itemize}
    \item \textbf{\textsc{Co-Matching} w/o Machine}. It only relies on the legal practitioner to identify the key sentences.
    \item \textbf{\textsc{Co-Matching} w/o Legal Practitioner}. It only relies on the matching machine to identify the key sentences.
    \item \textbf{\textsc{Co-Matching} w/ Intersection}. It is a heuristic method that takes the `\textit{intersection}` of the decisions of both parties. Specifically, if both parties have the same decisions on $d_i$, then the joint decision is in agreement with them; otherwise, $d_i$ is not key.
    \item \textbf{\textsc{Co-Matching} w/ Union}. It is a heuristic method that takes a `\textit{union}` of both decisions. Specifically, if either party considers $d_i$ is important, then $d_i$ is identified as a key sentence. If the two parties attribute different levels of importance to $d_i$, the approach selects the one with the higher level of importance. 
    \item \textbf{\textsc{Co-Matching} w/ Naive EM}. Unlike \textit{ProtoEM} which generates multiple confusion matrices for different prototypes, it generates only one confusion matrix using all historical documents.
\end{itemize}

\textbf{Dataset \& Evaluation Metrics}. Following our main baselines \cite{yu2022explainable, 10285038}, we experiment on ELAM and eCAIL~\cite{yu2022explainable}, two benchmark datasets with manually-labeled ground truth on key sentences. Specially, ELAM contains 1250 source legal cases, each associated with four target cases and four types of sentence labels. eCAIL is an extension of CAIL 2021 dataset\footnote{Fact Prediction Track data: \url{http://cail.cipsc.org.cn/}}. It contains 1875 source cases, each associated with three target cases and two types of sentence labels. Regarding the evaluation metrics, we also follow the previous studies \cite{shao2020bert, yu2022explainable, 10285038} and utilize the Accuracy, Precision, Recall, and F1 score. To evaluate the effectiveness of the human uncertainty estimation, we report the accuracy.

\textbf{Implementation details}. We conduct all our experiments using a single Nvidia RTX A6000, and we implement our codes in PyTorch. When it comes to machines, we use open-source checkpoints for experiments. If there are no available checkpoints, we train and reproduce those machines using open-source code from their official GitHub repository. We consider the manually labeled key sentences in the corresponding dataset as the ground truth. Additionally, we simulate the decisions of legal practitioners by adding random noise to the ground truth. According to \citet{shao2021investigating}, we simulate the decision accuracy of a legal practitioner with high expertise to be around 90\%\footnote{See the \textit{correct Cause per session} on the query formulation task in \citet{shao2021investigating}.} by adding 10\% noise to the ground truth. This is implemented by randomly choosing 10\% of the key sentences and marking them as 'Not Key'.
To understand the characteristics of \textsc{Co-Matching} when collaborating with legal practitioners with different domain expertise, 
we analyze the changes in decision accuracy among legal practitioners with varying domain expertise. This is done by modifying the noise rate, which ranges from 10\% to 50\%. The implementation of \textsc{Co-Matching} is flexible and can work with any matching machine capable of identifying sentence importance. In our experiments, we use {BERT-PLI} \cite{shao2020bert}, {IOT-Match} \cite{yu2022explainable}, and {GEIOT-Match} \cite{10285038} as machine matching baselines to collaborate with the legal practitioner. For the implementation of decision prototypes, we utilize K-Means (with $K=4$) to obtain the prototypes and use the EM algorithm to estimate the uncertainty, parameterized by the four confusion matrices. For all experiments, we consistently used 40 iterations for the \textit{ProtoEM} algorithm. For better understanding of the hyper-parameters of \textsc{Co-Matching}, we offer sensitivity analysis in Section \ref{thisanl}.

\subsection{Legal Case Matching Performance Evaluation}
\label{dyem}

\begin{table*}[]
\centering
\caption{Legal case matching performance evaluation. We mark our performance gain over corresponding methods using subscripts. \textsc{Co-Matching}, across different matching machines, constantly achieves significantly better performance than other baselines, which suggests that legal practitioners and their tacit knowledge are essential for the task of legal case matching. }
\label{tab:main}
\resizebox{0.98\textwidth}{!}{%
\begin{tabular}{l|cccc|cccc}
\toprule
\multicolumn{1}{c|}{\multirow{2}{*}{\textbf{\begin{tabular}[c]{@{}c@{}}Matching Methods\end{tabular}}}} & 
\multicolumn{4}{c|}{\textbf{eCAIL}} & \multicolumn{4}{c}{\textbf{ELAM}} \\ \cline{2-9}
\multicolumn{1}{c|}{} & \textbf{Acc.} & \textbf{Pre.} & \textbf{Rec.} & \textbf{F1} & \textbf{Acc.} & \textbf{Pre.} & \textbf{Rec.} & \textbf{F1} \\ \midrule
GPT-3.5-16K-turbo & 0.349  & 0.351 & 0.364 & 0.355 & 0.470 & 0.473 & 0.465 & 0.469 \\ \hline

*Sentence-BERT & 0.713 & 0.708 & 0.712 & 0.710 & 0.688 & 0.698 & 0.669 & 0.672 \\ 
*Lawformer &  0.707 & 0.702 & 0.706 & 0.699 & 0.699 & 0.723 & 0.683 & 0.692 \\
*Thematic Similarity$_{avg}$ &  0.715 & 0.709 & 0.713 & 0.710 & 0.710 & 0.713 & 0.690 & 0.691 \\
*Thematic Similarity$_{max}$ & 0.685 & 0.673 & 0.684 & 0.676 & 0.719 & 0.715 & 0.701 & 0.703 \\ 
\midrule

BERT-PLI & 0.701 & 0.704 & 0.712& 0.698 & 0.714 & 0.712 & 0.711 & 0.711\\
IOT-Match & 0.810 & 0.802 & 0.801 & 0.802 & 0.732 & 0.730 & 0.727 & 0.728 \\
GEIOT-Match  & 0.837 & 0.833 & 0.836  & 0.834 & 0.755 & 0.757 & 0.756 & 0.756 \\
\midrule

\textsc{Co-Matching$_{\text{BERT-PLI}}$}  & \textbf{0.752$_{+7.28\%}$} & \textbf{0.751$_{+6.68\%}$} & \textbf{0.750$_{+5.34\%}$} & \textbf{0.750$_{+7.45\%}$} & \textbf{0.760$_{+6.44\%}$}&\textbf{0.761$_{+6.88\%}$} &\textbf{0.763$_{+7.31\%}$} &\textbf{0.762$_{+7.17\%}$} \\ 
\textsc{Co-Matching$_{\text{IOT-Match}}$}  & \textbf{0.890$_{+9.88\%}$} & \textbf{0.893$_{+11.35\%}$} & \textbf{0.897$_{+11.99\%}$} & \textbf{0.895$_{+11.60\%}$} & \textbf{0.810$_{+10.66\%}$}  & \textbf{0.812$_{+11.23\%}$} & \textbf{0.811$_{+11.55\%}$} & \textbf{0.811$_{+11.40\%}$} \\ 
\textsc{Co-Matching$_{\text{GEIOT-Match}}$}  & \textbf{0.897$_{+7.17\%}$}&\textbf{0.899$_{+7.92\%}$}&\textbf{0.896$_{+7.18\%}$}&\textbf{0.897$_{+7.55\%}$}&\textbf{0.823$_{+9.01\%}$}&\textbf{0.822$_{+8.59\%}$}&\textbf{0.822$_{+8.73\%}$}&\textbf{0.822$_{+8.73\%}$} \\ \bottomrule
\textbf{Avg. Improvement}  & \textbf{+8.11\%} & \textbf{+8.65\%} & \textbf{+8.29\%}&\textbf{+8.87\%}& \textbf{+8.70\%}&\textbf{+8.90\%}& \textbf{+9.20\%} &\textbf{+9.10\%} \\ \bottomrule
\end{tabular}%
}
\end{table*}

This section presents a comparison of the \textsc{Co-Matching} with existing legal case matching baselines. The results are presented in Table \ref{tab:main}. Detailed observations are provided below.

\begin{itemize}
    \item \textbf{The performance of the GPT-3.5-16K-turbo is significantly inferior to that of the task-customized methods}. Although the existing legal-specific LLMs are not as effective as GPT-3.5-16K-turbo, as evidenced by previous benchmarks \cite{fei2023lawbench, dai2023laiw}, there is still a long way to go compared to methods tailored for legal case matching. The analysis of legal cases relies heavily on domain expertise and work experience in the judicial field. The lack of carefully crafted legal domain knowledge, such as tacit knowledge, for legal case matching significantly hinders the effectiveness of LLMs. This underscores the importance of human-machine collaboration for future study.
    \item \textbf{\textsc{Co-Matching} achieves superior results compared to baselines, suggesting that legal practitioners and their tacit knowledge are essential for the task of legal case matching}. As shown in Table \ref{tab:main}, \textsc{Co-Matching} significantly enhances the matching performance compared to all baselines. This improvement is consistent across all evaluation dimensions. When averaged across all metrics and datasets, the performance of \textsc{Co-Matching} has shown a 6.82\% increase compared to BERT-PLI, an 11.21\% increase compared to IOT-Match, and an 8.11\% increase compared to the SOTA method, GEIOT-Match. These results indicate that 1) \textsc{Co-Matching} is independent of the matching machine used for key sentence identification. 2) Additionally, they underscore the significance of involving legal practitioners in the process of legal case matching. The tacit knowledge of legal practitioners, while hard to be articulated into the machine, is crucial for enhancing the matching performance.
\end{itemize}

\begin{table*}[]
\centering
\caption{Evaluation on human-machine collaboration and uncertainty estimation. \textsc{Co-Matching} outperforms both human- and machine-based performance in isolation. It also guarantees improved effectiveness in human-machine collaboration. Finally, finer-grained uncertainty estimation via \textit{ProtoEM} brings superior matching performance compared to using Naive EM. }
\label{tab:main2}
\resizebox{0.98\textwidth}{!}{%
\begin{tabular}{l|l|cccc|cccc}
\toprule
\multicolumn{1}{c|}{\multirow{2}{*}{\textbf{\begin{tabular}[c]{@{}c@{}}Machine\end{tabular}}}} & \multicolumn{1}{c|}{\multirow{2}{*}{\textbf{\begin{tabular}[c]{@{}c@{}}Key Sentence Identification\end{tabular}}}} & \multicolumn{4}{c|}{\textbf{eCAIL}} & \multicolumn{4}{c}{\textbf{ELAM}} \\ \cline{3-10}
\multicolumn{1}{c|}{} & \multicolumn{1}{c|}{} & \textbf{Acc.} & \textbf{Pre.} & \textbf{Rec.} & \textbf{F1} & \textbf{Acc.} & \textbf{Pre.} & \textbf{Rec.} & \textbf{F1} \\ \midrule

\multirow{6}{*}{BERT-PLI} & \textsc{Co-Matching} & \textbf{0.752} & \textbf{0.751} & \textbf{0.750} & \textbf{0.750} & \textbf{0.760} & \textbf{0.761} & \textbf{0.763} &\textbf{0.762} \\ \cline{2-10}
& \textsc{Co-Matching} w/o Legal Practitioner & 0.701 & 0.704 & 0.712& 0.698 & 0.714 & 0.712 & 0.711 & 0.711\\
& \textsc{Co-Matching} w/o Machine & 0.721 & 0.717 & 0.718 & 0.717 & 0.730 & 0.731&0.730 &0.730 \\ \cline{2-10}
& \textsc{Co-Matching} w/ Naive EM  & 0.733 & 0.735 & 0.734 & 0.734 & 0.744& 0.747& 0.748& 0.747\\
& \textsc{Co-Matching} w/ Intersection  & 0.725
 & 0.720 & 0.722 & 0.721 & 0.730 & 0.733 & 0.734 & 0.733\\
& \textsc{Co-Matching} w/ Union  & 0.712 & 0.713 & 0.715 & 0.714 & 0.720 & 0.722 & 0.719 & 0.720 \\ \midrule

\multirow{6}{*}{IOT-Match} & \textsc{Co-Matching}  & \textbf{0.890} & \textbf{0.893} & \textbf{0.897} & \textbf{0.895} & \textbf{0.810}  & \textbf{0.812} & \textbf{0.811} & \textbf{0.811} \\ \cline{2-10}
& \textsc{Co-Matching} w/o Legal Practitioner & 0.810 & 0.802 & 0.801 & 0.802 & 0.732 & 0.730 & 0.727 & 0.728 \\
& \textsc{Co-Matching} w/o Machine & 0.847 & 0.833 & 0.840 & 0.836 & 0.767 & 0.769  & 0.773 & 0.771 \\ \cline{2-10}
& \textsc{Co-Matching} w/ Naive EM &0.857 & 0.851 & 0.848 & 0.849 & 0.788 & 0.784 & 0.787 & 0.785 \\
& \textsc{Co-Matching} w/ Intersection  & 0.835 & 0.833   & 0.832  & 0.832 & 0.757 & 0.758 & 0.758 & 0.757 \\ 
& \textsc{Co-Matching} w/ Union  & 0.752  & 0.743 & 0.741 & 0.742 & 0.767 & 0.760 & 0.765 & 0.762 \\
\midrule

\multirow{6}{*}{GEIOT-Match} & \textsc{Co-Matching} &  \textbf{0.897}&\textbf{0.899}&\textbf{0.896}&\textbf{0.897}&\textbf{0.823}&\textbf{0.822}&\textbf{0.822}&\textbf{0.822} \\ \cline{2-10}
& \textsc{Co-Matching} w/o Legal Practitioner  & 0.837 & 0.833 & 0.836  & 0.834 & 0.755 & 0.757 & 0.756 & 0.756 \\
& \textsc{Co-Matching} w/o Machine & 0.842&0.845 &0.844 &0.844&0.773&0.775&0.777& 0.776\\ \cline{2-10}
& \textsc{Co-Matching} w/ Naive EM  & 0.860& 0.861& 0.862& 0.861 &0.793& 0.791 &0.790& 0.790 \\
& \textsc{Co-Matching} w/ Intersection  &  0.830&0.831&0.834&0.832&0.778&0.782&0.784&0.783\\
& \textsc{Co-Matching} w/ Union  & 0.778&0.772&0.773&0.772&0.761&0.765&0.764&0.764\\ 
\bottomrule

\end{tabular}%
}
\end{table*}

\subsection{In-depth Analysis on \textsc{Co-Matching}}
\label{dyem2}
We analyze the advantages and characteristics of \textsc{Co-Matching}, which achieves human-machine joint decision-making on key sentences in a probabilistic manner and involves the \textit{ProtoEM} to estimate human decision uncertainty without assessing the ground truth. 
In this section, we delve into the human-machine collaboration characteristics of \textsc{Co-Matching} in Section \ref{this2}. Moreover, we investigate the potential of \textsc{Co-Matching} for collaboration with legal practitioners of varying levels of tacit knowledge in Section \ref{this3}. Finally, we step into the human uncertainty estimation of \textsc{Co-Matching}, revealing its impact on collaborative performance in Section \ref{dyem3}.
Overall, our findings indicate that \textsc{Co-Matching} establishes successful collaborative matching between the legal practitioner and the machine, resulting in superior matching results. Detailed observations are provided below.

\begin{figure*}
    \centering
    \includegraphics[width=0.999\textwidth]{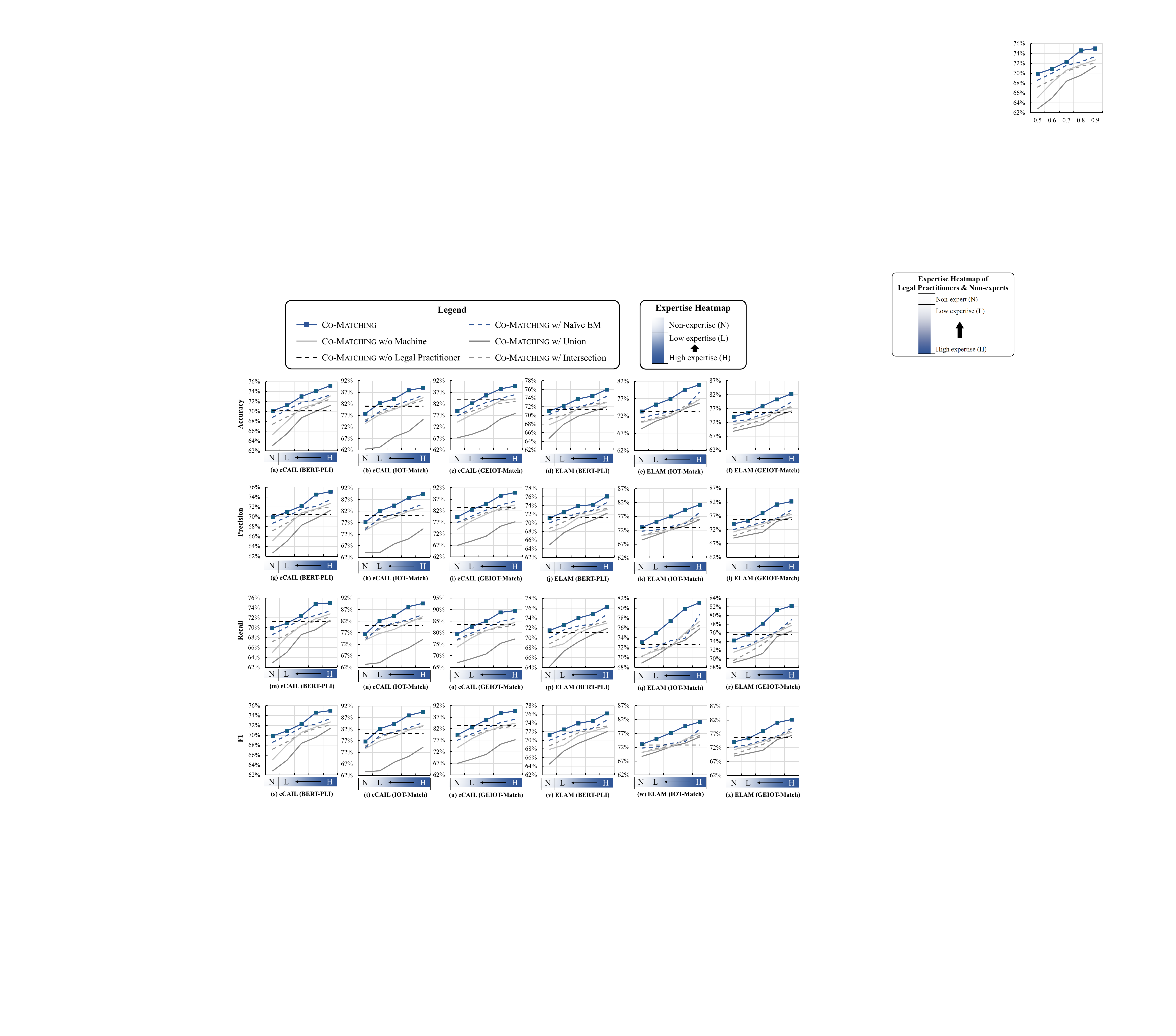}
    \caption{Matching performance when collaborating with legal practitioners of varying levels of tacit knowledge.
    \textsc{Co-Matching} has strong adaptability to different legal practitioners compared to other baselines. Non-experts hinder the realization of human-machine complementarity due to their lack of legal tacit knowledge.}
    \label{fig:noise}
\end{figure*}

\subsubsection{Human-machine collaboration characteristics}
\label{this2}
In this section, we explore two types of ablation baselines, with the aim to sort out the performance variation of different methods: The first one separates the impact of the legal practitioner and the machine, namely, \textit{\textsc{Co-Matching} w/o Legal Practitioner} and \textit{\textsc{Co-Matching} w/o Machine}. The second one explores alternative methods for integrating human- and machine-selected key sentences, including \textit{\textsc{Co-Matching} w/ Intersection} and \textit{\textsc{Co-Matching} w/ Union}. 

\begin{itemize}
    \item \textbf{\textsc{Co-Matching} promotes the matching performance, outperforming both human- and machine-based matching performance in isolation}. The results in Table \ref{tab:main2} demonstrate that \textsc{Co-Matching} consistently outperforms \textsc{Co-Matching} w/o Legal Practitioner and \textsc{Co-Matching} w/o Machine. Specifically, when averaged across different machine backbones, datasets, and evaluation metrics, \textsc{Co-Matching} surpasses the legal practitioner by 5.51\% and the machine by 8.71\%. This improvement highlights the potential for collaboration between the legal practitioner and the machine in identifying and selecting key sentences, leading to better legal case matching results. One possible reason for such improvement is that the legal practitioner and the machine may have access to both shared and non-overlapping information. For example, a judge learns about the predisposition of the defendant through interaction, while a machine extracts complex features of a legal case via tons of texts and documents. This scenario may lead to suboptimal decision-making strategies by both the legal practitioner and the machine. In response, \textsc{Co-Matching} provides a solution for building a human-machine collaborative matching, which promotes information complementarity.
    \item \textbf{\textsc{Co-Matching} ensures better human-machine collaboration effectiveness}. When averaged across different machine backbones, datasets, and evaluation metrics, \textsc{Co-Matching} surpasses the Intersection-based method by 5.89\%, the Union-based method by 10.15\%, as well as the Naive EM by 3.44\%. The effectiveness of \textsc{Co-Matching} compared to these heuristic ablation baselines, which do not account for decision uncertainty, suggests that explicitly modeling human uncertainty may be a more effective approach to support human-machine collaboration and joint decision-making. To sum up, uncertainty often hinders the decision-making process and task execution, making uncertainty modeling a significant consideration, as emphasized by our findings.
\end{itemize}

\begin{figure}[h]
    \centering
\includegraphics[width=0.999\textwidth]{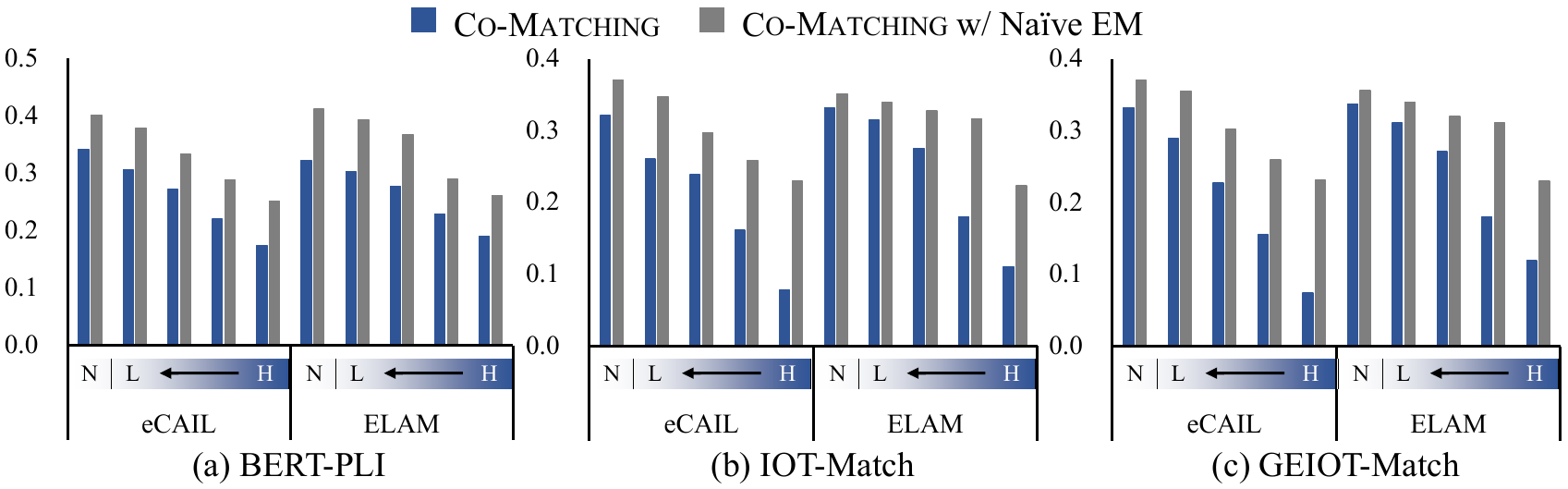}
    \caption{Illustration on uncertainty estimation error. \textsc{Co-Matching} enjoys lower estimation error, thanks to the finer-grained uncertainty estimation via \textit{ProtoEM}.}
    \label{fig:error}
\end{figure}

\subsubsection{Collaborating with legal practitioners of varying levels of tacit knowledge}
\label{this3}
In real-world scenarios, there are significant differences in legal matching accuracy among different domain expertise groups \cite{10.1145/3569929}. Notably, individuals without a legal background make a significantly higher number of errors compared to those with varying levels of tacit knowledge. We explore the potential collaboration of \textsc{Co-Matching} with legal practitioners possessing varying levels of tacit knowledge. We adjust the noise rate in the decisions made by legal practitioners, ranging from 10\% to 50\%, where 50\% noise represents Non-Experts making random decisions and selecting sentences due to a lack of implicit domain knowledge. The results are illustrated in Fig. \ref{fig:noise}. Detailed observations are provided below.

\begin{itemize}
    \item \textbf{Compared to others, \textsc{Co-Matching} is better suited for legal practitioners of varying levels of tacit knowledge}. We discover that the performance of all methods is influenced by the tacit knowledge level of legal practitioners, with lower tacit knowledge levels leading to diminished matching performance. When involving less experienced legal practitioners, most human-machine collaboration methods struggle to maintain the complementarity between the legal practitioners and the machine, resulting in their effectiveness being inferior to that of tasks conducted by the legal practitioners or the machine alone. However, we observe that \textsc{Co-Matching} continues to deliver strong matching performance, surpassing both human- and machine-based matching performance in isolation. We attribute this improvement in performance to the finer-grained estimation of human uncertainty, which allows us to "filter out" low-quality human decisions in a probabilistic way. This highlights the importance of incorporating human uncertainty estimation into the human-machine collaboration process to avoid the issue of "over-trust" in humans \cite{cha2021human}.
    \item \textbf{Non-experts are not suitable for participating in legal case matching tasks, hindering the realization of human-machine complementarity}. Legal case matching demands a profound understanding of the law, legal principles, and case precedents. Non-experts (i.e., 50\% noise rate) may lack the expertise needed to accurately interpret and identify key sentences in legal documents. Let alone the tacit knowledge for complex and nuanced analysis, which necessitates specialized training and experience. As a result, their involvement in legal case search tasks leads to inaccurate results, and they should not participate in legal case matching, considering the justice of legal proceedings. Moreover, despite the fact that our research focuses on the critical role of the legal practitioner, our experiment results also reveal, to some extent, the importance of the machine. Specifically, the effectiveness of \textsc{Co-Matching} also improves with the improvement of the machine's performance (cf. Table \ref{tab:main} and Fig.\ref{fig:noise}). However, a poorly performing machine may also be unable to effectively collaborate with the legal practitioner. In fact, such poorly performing machines would not be deployed, especially in high-stakes fields such as legal case matching. From a broader perspective, the future development of human-machine collaboration should focus more on assessing the strengths and capabilities of each collaborator and selecting the appropriate collaborators.
\end{itemize}

\begin{figure}
    \centering
\includegraphics[width=0.999\textwidth]{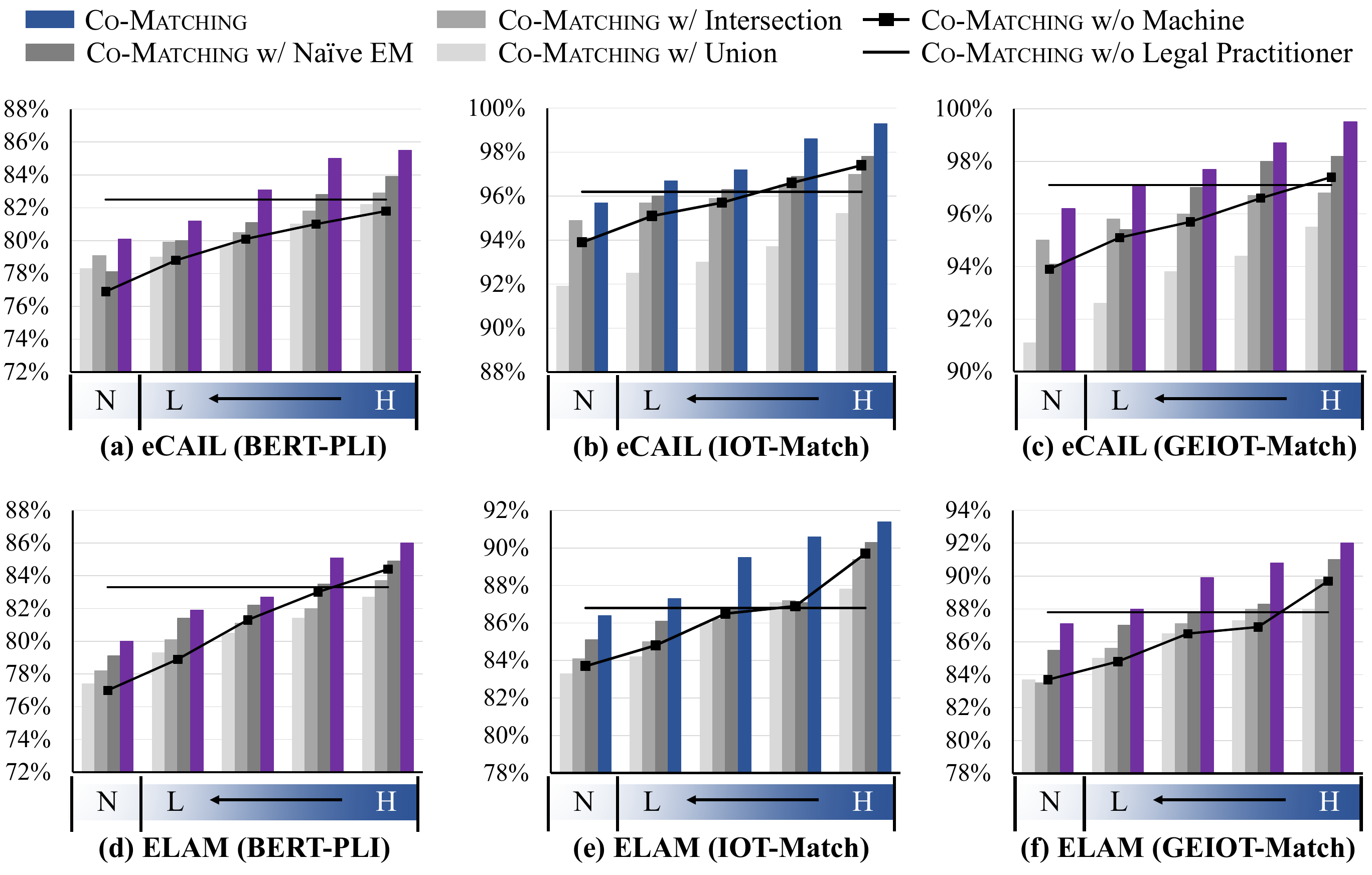}
    \caption{Illustration on key sentence identification accuracy. \textsc{Co-Matching} enjoys higher accuracy than other human-machine based baselines.}
    \label{fig:acc}
\end{figure}

\subsubsection{Uncertainty estimation characteristics}
\label{dyem3}
\textsc{Co-Matching} employs \textit{ProtoEM} to estimate the human uncertainty at the prototype level, while \textsc{Co-Matching} w/ Naive EM utilizes all historical documents for estimation. In this section, we aim to explore and understand the characteristics of uncertainty estimation. The experimental results are illustrated in Table \ref{tab:main2}, Fig. \ref{fig:error}, and Fig. \ref{fig:acc}. Detailed observations are provided below.

\textbf{Finer-grained uncertainty estimation via \textit{ProtoEM} brings more accuracy on key sentence identification and secures our superiority in matching performance}. As illustrated in Table \ref{tab:main2}, \textsc{Co-Matching} surpasses the Naive EM based method by 3.83\% on eCAIL and 3.05\% on ELAM. Naive EM is criticized for being too coarse-grained to capture subtle human decision-making behaviors, leading to inaccurate estimation. Instead, \textit{ProtoEM} enables the estimation of human uncertainty in a finer-grained manner by introducing multiple decision prototypes and utilizing the most relevant decision prototype to quantify the uncertainty of human decisions during collaborative matching. 
For comprehensive understanding, we present a detailed analysis to uncover the reasons for the superiority of \textit{ProtoEM}. This includes the analysis regarding the uncertainty estimation error and key sentence identification accuracy.
\begin{itemize}
    \item \textbf{Error analysis on uncertainty estimation}. We demonstrate the errors in uncertainty estimation for both \textsc{Co-Matching} and its variance using Naive EM in Fig.\ref{fig:error}, where the error is quantified by the average Frobenius Norm, $\sum_{d_i \in D} \| \phi_i - \phi_i^0 \|_F / |D|$. Here, $\phi_i^0$ is the ground truth confusion matrix calculated using the ground truth labels contained in the each dataset. As illustrated in Fig.\ref{fig:error}, our \textit{ProtoEM} has demonstrated a 22.42\% improvement on BERT-PLI, a 28.41\% improvement on IOT-Match, and a 28.14\% improvement on GEIOT-Match when compared to Naive EM, across all datasets. These findings suggest that, in comparison to Naive EM, the uncertainty estimated by \textit{ProtoEM} consistently aligns closer to the ground truth when collaborating with legal practitioners possessing varying levels of tacit knowledge. This improved estimation of human decision-making uncertainty lays a stronger groundwork for generating joint human-machine decisions on each sentence $d_i$, which contributes to our superior performance.
    \item \textbf{Accuracy analysis on key sentence identification}. Clearly, reduced uncertainty estimation errors contribute to improved performance in human-machine joint decision-making, leading to more accurate identification of key sentences. For better understanding, we present the accuracy on key sentences of different methods. As shown in Fig.\ref{fig:acc}, the results indicate the superiority of our \textsc{Co-Matching}, which has demonstrated a 1.67\% improvement on BERT-PLI, a 1.80\% improvement on IOT-Match, and a 1.61\% improvement on GEIOT-Match when compared to \textsc{Co-Matching} w/ Naive EM, across all datasets. When working with non-experts who have minimal legal tacit knowledge, all forms of human-machine collaboration show reduced effectiveness compared to \textit{\textsc{Co-Matching} w/o Legal Practitioner}. Nevertheless, our \textsc{Co-Matching} does manage to achieve performance closest to \textit{\textsc{Co-Matching} w/o Legal Practitioner}. This suggests that upon recognizing the high uncertainty in human decision-making, \textsc{Co-Matching} may exhibit a tendency to place greater trust in machine decision-making to some extent. Nonetheless, we emphasize that non-experts are not suitable for participating in high-stakes legal case matching tasks, which hinders the realization of human-machine complementarity, as discussed in Section \ref{this3}.

\end{itemize}

To sum up, in the context of human-machine collaboration, employing fine-grained uncertainty modeling in human decision-making enhances the effectiveness of the collaboration.


\subsection{Sensitivity Analysis}
\label{thisanl}
In this section, we investigate the sensitivity of the hyperparameters in \textsc{Co-Matching}, specifically the number of K-means clusters used to derive the prototypes and the number of EM iterations in the \textit{ProtoEM}. It's important to note that these two hyperparameters impact our \textsc{Co-Matching} through their influence on the \textit{ProtoEM} algorithm and no other hyperparameters are involved in our \textsc{Co-Matching}.

\begin{figure*}
    \centering
\includegraphics[width=0.999\textwidth]{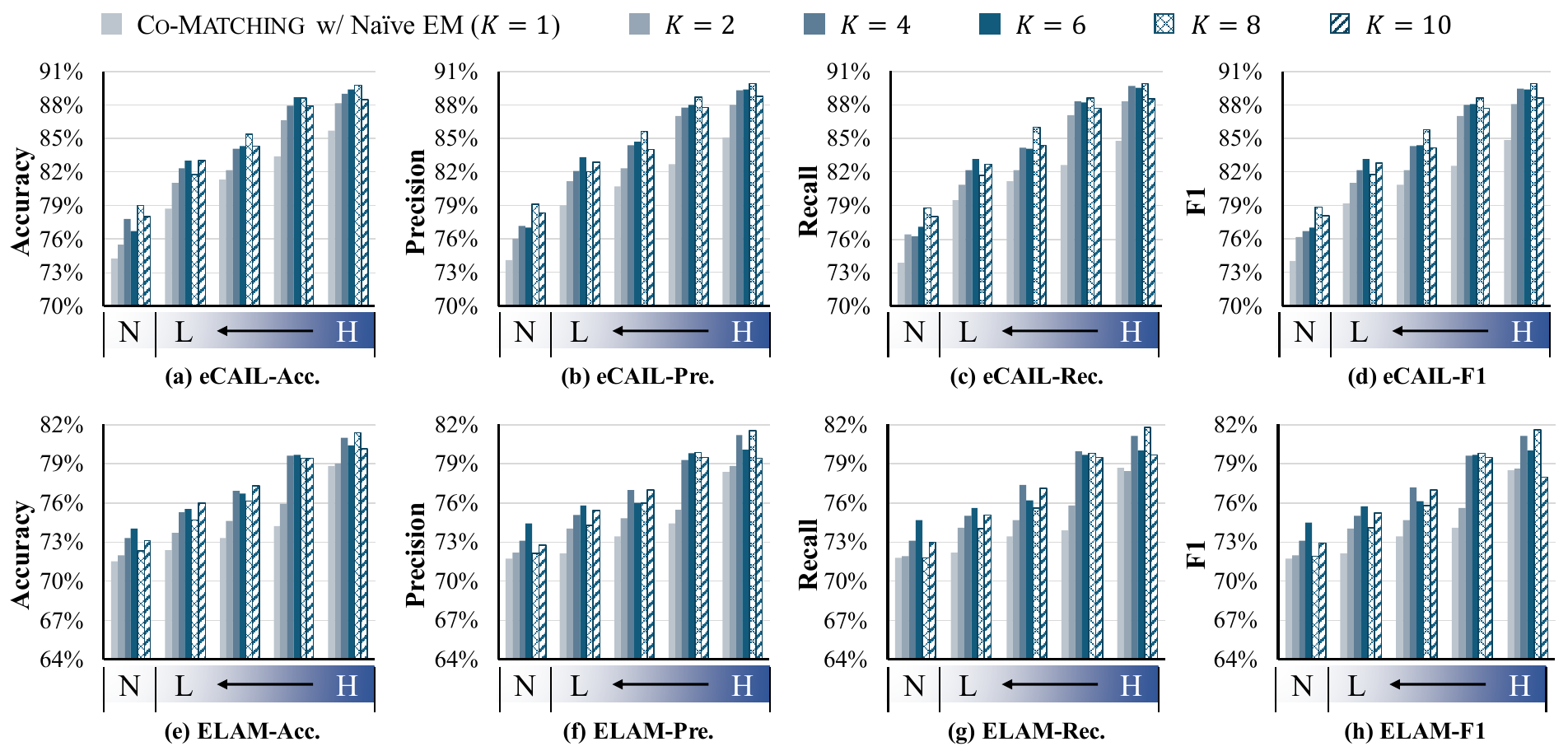}
    \caption{Sensitivity analysis on the number of K-means clusters, varying in $K=\{1,2,4,6,8,10\}$. \textsc{Co-Matching} consistently surpasses the \textit{\textsc{Co-Matching} w/ Naive EM} across various settings of the hyperparameter $K$. However, it's important to note that simply increasing $K$ doesn't necessarily lead to improved performance. $K=4, 6$, or $8$ empirically yields the better results. We set $K=4$ in our main experiments.}
    \label{fig:sensitive-kmeans}
\end{figure*}

\subsubsection{Sensitivity on the Number of K-means Clusters}
In Section \ref{asdrfwef}, we quantify the uncertainty of the human decision by building analogies to the uncertainty of the most relevant decision prototype. This process involves clustering the historical data into prototypes, which is accomplished using the K-Means algorithm in our experiments. We adjust the number of K-means clusters $K$ within the set $\{1, 2, 4, 6, 8, 10\}$ and assess the performance of \textsc{Co-Matching}. It's important to note that when $K=1$, \textsc{Co-Matching} transforms into \textit{\textsc{Co-Matching} w/ Naive EM}. Taking the IOT-Match as the matching method, Fig.\ref{fig:sensitive-kmeans} visually presents the experimental results, leading us to the following observations.

\begin{itemize}
    \item \textbf{Across all the configurations on the number of K-means clusters, \textsc{Co-Matching} consistently outperforms \textit{\textsc{Co-Matching} w/ Naive EM}}. More precisely, \textsc{Co-Matching} demonstrates a consistent average improvement of 4.26\% compared to \textit{\textsc{Co-Matching} w/ Naive EM} when considering various combinations of $K$ values ($K \geq 2$) and tacit knowledge levels. This result clearly demonstrates the value of incorporating fine-grained uncertainty estimations into models, leading to more robust and accurate performance.
    \item \textbf{Simply increasing $K$ does not necessarily lead to improved performance}. Our findings show that increasing the number of clusters (i.e., $K > 1$) leads to a noticeable performance gain compared to \textit{\textsc{Co-Matching} w/ Naive EM} (i.e., $K=1$). This is because using a finer-grained approach with more clusters allows for a more accurate estimation of user decision uncertainty. However, we also observed that pushing this too far, such as setting $K=10$, can actually hinder performance. This decline likely stems from having fewer sentences within each cluster as the number of clusters grows, which in turn reduces the effectiveness of the \textit{ProtoEM} algorithm's estimations. This effect is particularly pronounced when dealing with lower levels of tacit knowledge, for instance, when collaborating with non-experts. Therefore, in our main experiments, we set $K=4$ for all matching methods and datasets without further parameter tuning.
\end{itemize}

\begin{figure*}
    \centering
\includegraphics[width=0.999\textwidth]{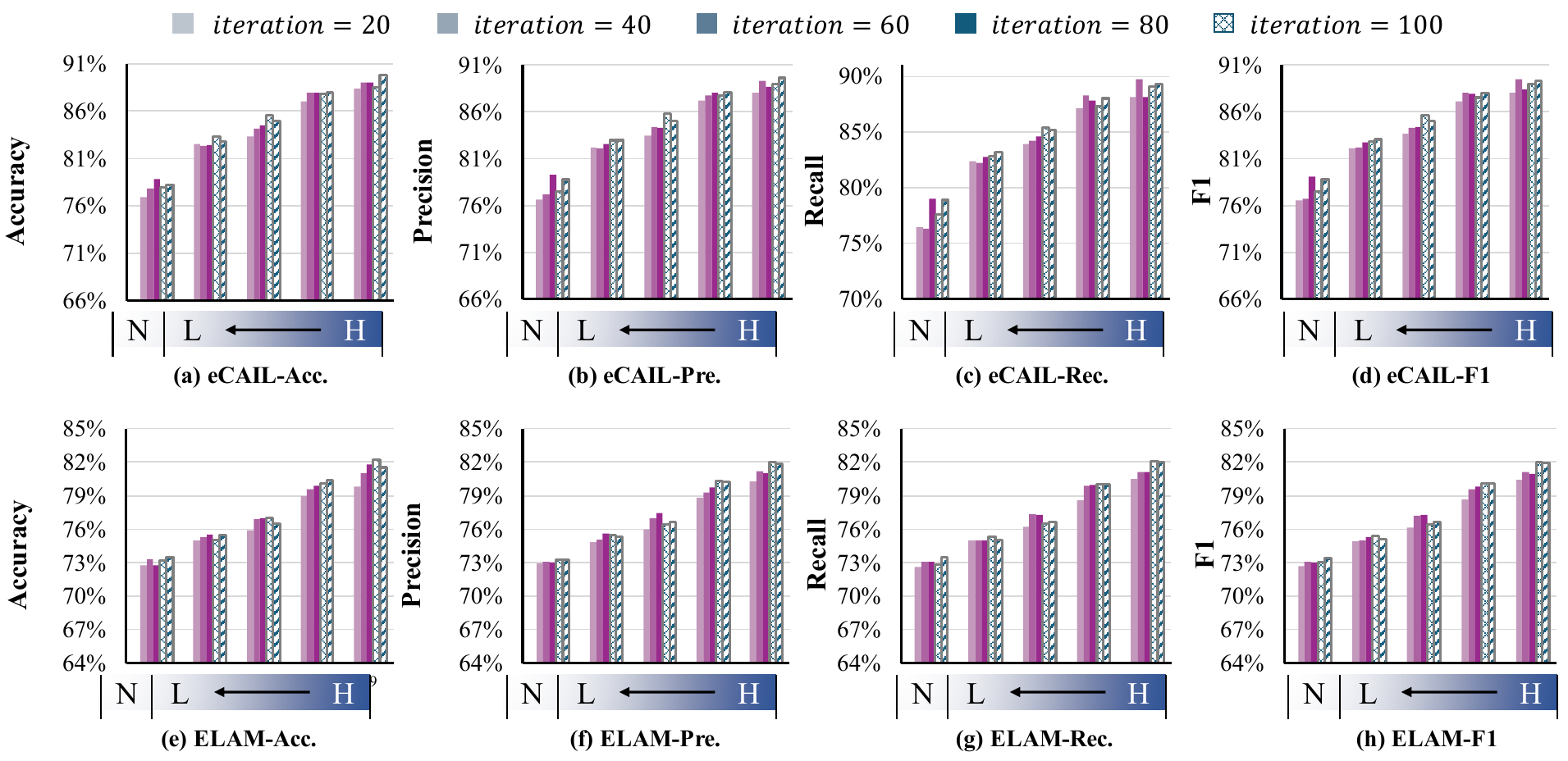}
    \caption{Sensitivity analysis on the number of EM iterations, tuning within a specific range $\{20, 40, 60, 80, 100\}$. Increasing the number of iterations in the \textit{ProtoEM} algorithm may lead to improved performance> However, occasional exceptions might occur due to overfitting. Therefore, for all experiments, we consistently used 40 iterations for the \textit{ProtoEM} algorithm.}
    \label{fig:sensitive-em}
\end{figure*}

\subsubsection{Sensitivity on the Number of EM Iterations in \textit{ProtoEM}}
In Section \ref{emduwe}, the \textit{ProtoEM} method employs the EM algorithm to estimate the uncertainty of decisions for each prototype. This entails iteratively executing the E-step and the M-step until convergence is achieved. To investigate its impact, we vary the number of iterations within a specific range $\{20, 40, 60, 80, 100\}$ and assess the performance of \textsc{Co-Matching}. Taking the IOT-Match as the matching method, we illustrate the experimental results in Fig.\ref{fig:sensitive-em}. Detailed observations are as follows.

\begin{itemize}
    \item \textbf{Increasing the number of iterations in the \textit{ProtoEM} algorithm may lead to improved performance, but not always}. In general, both the EM and \textit{ProtoEM} algorithms require a specific number of iterations to converge. As Fig.\ref{fig:sensitive-em} demonstrates, \textsc{Co-Matching} generally exhibits improved performance with increasing iterations. This trend holds true for legal practitioners with diverse levels of tacit knowledge. While occasional exceptions might occur where excessive iterations negatively impact \textsc{Co-Matching}'s effectiveness, potentially leading to overfitting in the EM calculations, we opted for 40 iterations in our main experiment. This choice enjoys the computational efficiency and good performance, minimizing the cost of excessive iterations without significantly affecting the results.
\end{itemize}

To sum up, our experiments reveal that blindly increasing the number of clusters in K-means or the number of iterations in EM algorithms does not guarantee better performance. While these hyperparameters can influence the outcome, their effect is often minor and may not significantly impact the results.


\section{Conclusion}
Legal case matching necessitates the tacit knowledge of legal practitioners, which is difficult to specify for encoding into the machine. In this paper, we emphasize the significance of maintaining the legal practitioner’s participation to deliver promising performance, and the potential for collaboration between the legal practitioner and the machine to work as a team.
In light of this, we pioneer work on human-machine collaborative matching and tentatively provide a practical framework, \textsc{Co-Matching}.
\textsc{Co-Matching} involves both the legal practitioner and the machine making decisions, incorporating tacit knowledge. 
\textsc{Co-Matching} also incorporates a novel method called \textit{ProtoEM} to estimate human decision uncertainty, which facilitates the joint decision-making of discrete human decisions and probabilistic machine decisions with no ground truth.
We experimentally show that \textsc{Co-Matching} consistently outperforms both human- and machine-based matching performance in isolation. Additionally, it collaborates well with legal practitioners possessing varying levels of tacit knowledge. 

Recent advancements in deep learning have led to the development of machines that can handle complex tasks independently and automatically. This increased level of autonomy is reshaping the way humans interact with machines, moving away from meticulous control towards higher-level collaborations. Our research represents a groundbreaking effort in the area of human-machine collaboration for the matching task. 
Moreover, our human-machine collaboration paradigm also surpasses cooperation and entails a more integrated and interactive approach to achieving a shared and coordinated goal, where human and machine collaborate with each other in an equal partnership, marking a milestone for future collaborative matching studies. 
Looking ahead, we plan to extend our framework into the conversational legal case matching setting, allowing legal professionals to engage in multi-turn conversations with the matching machine. To achieve this, the \textit{ProtoEM} used in the \textsc{Co-Matching} could be extended to an online version with sequential EM updates at the end of each conversation.
\clearpage

\bibliographystyle{ACM-Reference-Format}
\bibliography{sample-base}

\end{document}